\documentclass[%
 reprint,
 amsmath,amssymb,
 aps, prl
]{revtex4-1}

\usepackage{graphicx}% Include figure files
\usepackage{dcolumn}% Align table columns on decimal point
\usepackage{bm}% bold math
\usepackage{amsthm}
\usepackage{amsfonts}

\usepackage{commath}

\usepackage{float}

\begin{document}

\preprint{APS/123-QED}

\title{Graphene Terahertz Absorption}% Force line breaks with \\

%\affiliation{%


%\affiliation{
%\affiliation{
 %Third institution, the second for Charlie Author
%\affiliation{%


\author{Yuan Yang}
\affiliation{Department of Chemistry and Chemical Biology, Harvard University, Cambridge, MA, 02138, United States}
\author{Grigory Kolesov}
\affiliation{John Paulson School of Engineering and Applied Sciences, Harvard University, Cambridge, MA, 02138, United States}
\author{Lucas Kocia}
\affiliation{Department of Physics, Tufts University, Medford, MA, 02155, United States}
\author{Eric J. Heller}
%\homepage{http://www-heller.harvard.edu/}
\email{heller@physics.harvard.edu}
\affiliation{Department of Physics, Harvard University, Cambridge, MA, 02138, United States}
\affiliation{Department of Chemistry and Chemical Biology, Harvard University, Cambridge, MA, 02138, United States}


\begin{abstract}
The unique terahertz properties of graphene has been identified for novel optoelectronic applications. In a graphene sample with bias voltage added, there is an enhanced absorption in the far infrared region and a diminished absorption in the infrared region. The strength of enhancement(diminishment) increases with the gate voltage, and the enhancement compensates the diminishment. We find that it is the coherence length of electrons in graphene that allows pure electronic transitions between states differing by small momentums and makes intraband transition possible, is responsible for the far infrared enhancement. Phonon assisted processes are not necessary and would not in any case contribute to a sum rule. This naturally leads to results obeying the general sum-rule in optical absorptions. Our prediction  of the strength of enhancement(diminishment) in terms of the bias agrees with experiments.  This is the first direct calculation we are aware of, since the prior phonon assisted model for indirect transition should not obey a sum rule.
%\begin{description}
%\end{description}
\end{abstract}

\maketitle

%\tableofcontents
\section{INTRODUCTION}

Graphene provides a unique material system to study Dirac fermion physics in two dimensions. Researchers have demonstrated exotic Dirac fermion phenomena  in graphene, ranging from anomalous quantum Hall effects\cite{Novoselov2005, Zhang2005} to Klein tunneling in low-frequency (DC) electrical transport\cite{Katsnelson2006}. They also observed an optical conductance defined by the fine-structure constant\cite{PhysRevLett.101.196405, Nair1308} and gate-tunable infrared (IR) absorption in Dirac fermion interband transitions\cite{Li2008, Wang206}. Situated between DC electrical transport and interband optical excitation is the spectral range dominated by intraband transitions. This intraband dynamics response has attracted much recent attention and  is expected to play a key role in the future development of ultrahigh-speed electronics at terahertz (THz) frequencies and THz-to-mid-IR optoelectronic devices\cite{PhysRevLett.95.187403, Li2008, doi:10.1021/nl301496r, doi:10.1021/acs.nanolett.6b00405, Mak20121341, Gan2013, Liu2011, 1882-0786-7-11-115101, Zhang2015}. 

\begin{figure}[h]
\includegraphics[width=9cm]{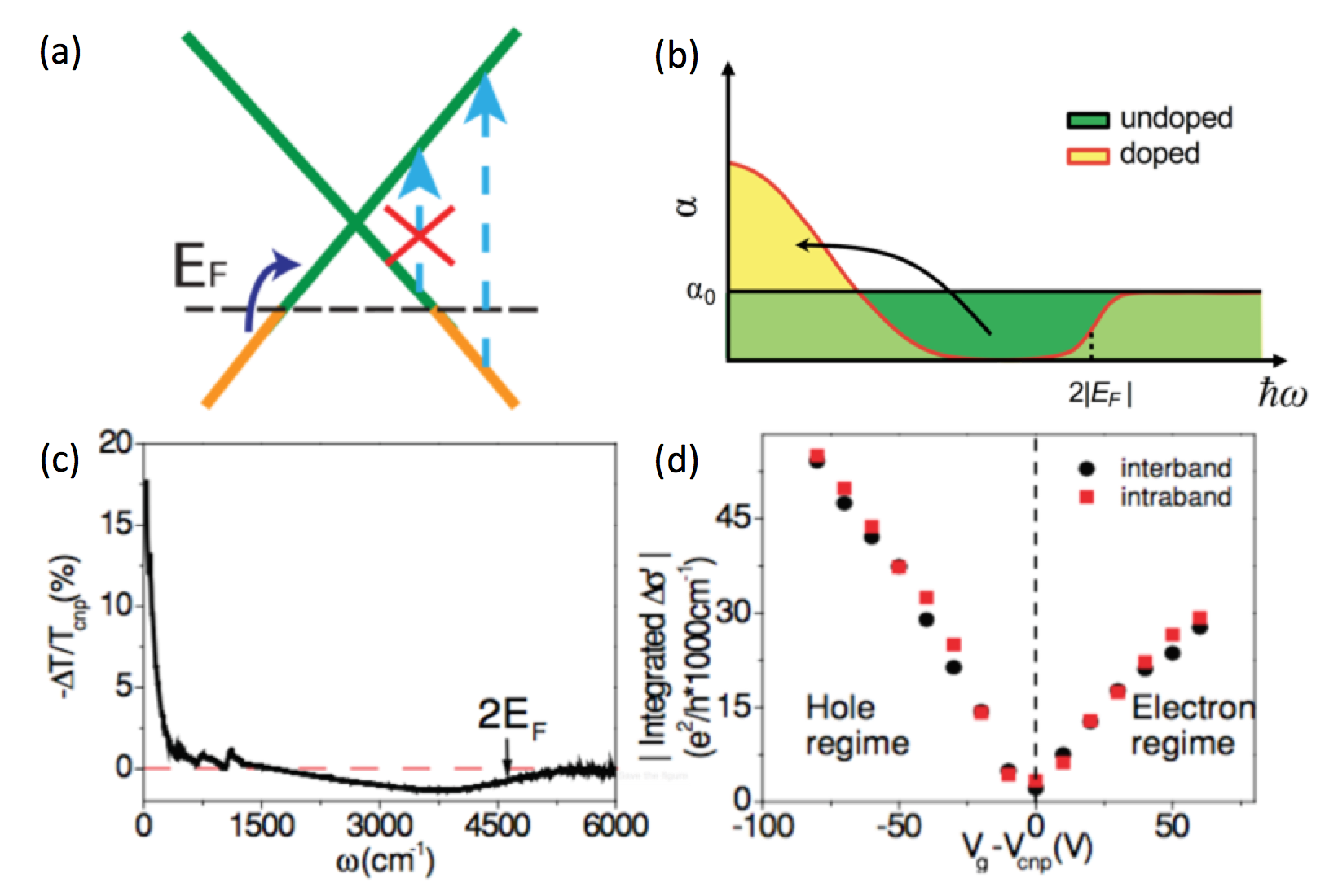}
\caption{Graphene far infrared and infrared region absorption: (a) is an illustration of intraband(solid arrow) and interband (dash arrow) transitions in hole-doped graphene. (b) is an illustration of the corresponding absorption
spectrum for the doped (yellow) and undoped (green) case. When graphene is doped, the   absorption strength below $2\abs{E_F}$ is transferred to lower energy(black arrow). (c) shows the experimental gate-induced change of IR transmittance $\Delta T/T$ through graphene at Vg =-70 V (hole doped) compared to transmittance at the charge neutral graphene. The spectrum shows an increase of absorption at low wave numbers in the far infrared region and a reduction of absorption at higher wave numbers in the infrared region. (d) shows the experimental integrated value of the enhancement and diminishment in optical absorption as a function of gating voltage. The change of the interband contribution is equal to that of the intraband parts. Intraband absorption increases with carrier doping, while interband transitions up to $2\abs{E_F}$ become forbidden due to empty initial states. Figures (a), (c) and (d) are adapted from the work of Hornig etc,\cite{horng2011drude} and figure (b) is adapted from the work of Kim etc.\cite{Kim2013}}
\label{terahertz_exp}
\end{figure}

In a tunable carrier concentration graphene sample, intraband absorption increases as the Fermi energy, $E_f$, moves away from the Dirac point in either direction(p-type or n-type). On the other hand, interband absorption is possible only when the photon energy is larger than $2E_F$, see figure~\ref{terahertz_exp}(a)(b)(c). In the absorption curve, the rise in the far infrared region compared to the universal absorption compensates the dip in the infrared region, figure~\ref{terahertz_exp}(d)\cite{horng2011drude, Kim2013}. The capability of tuning the type and concentrations of charge carriers and conductivity in graphene is desired for many electronic and optoelectronic applications. It is essential to understand the true mechanism of the far-infrared and infrared transitions in order to make use of them in tuning the electronic properties in graphene devices.

The Drude model is used to understand the optical conductivity of graphene\cite{PhysRevLett.96.256802, horng2011drude, RevModPhys.82.2673, Kim2013}. In the Drude model, electrons are treated as classical charged particles moving under an electric field. It neglects any long-range interaction between the electron and the ions or between the electrons. The only possible interaction of a free electron with its environment is via instantaneous collisions. The quantum dynamics under the model is blurred. In this work, we develop a quantum description of both intraband and interband transitions. We show that it is the coherence length of electron that makes the intraband transition in a doped graphene sample possible. Both intraband and interband transitions are due to the electronic excitations between quantum states. This naturally leads that the enhancement in the far-infrared region compensates the diminishment in the infrared region, which gives the result obeying the sum-rule.

 Phonon assisted processes have been invoked to account for the intraband component\cite{Mak20121341}.  Phonons seem necessary when thinking of strict momentum conservation for a non-vertical intraband transition (figure~\ref{terahertz_exp}(a)).  However phonon assisted processed cannot in any case contribute to the desired sum rule, since the other, compensating direct interband transitions do not involve a phonon.    A sum rule cannot involve physically different processes with unrelated terms   in their matrix elements.  Since Bloch wave momentum is less than perfectly definite due to finite coherence lengths, momentum conservation is only approximately enforced by  Bloch waves. Slightly momentum non-conserving processes may be drawn in the usual band structure diagrams.

\section{Results and Discussion}

The optical absorption of graphene in the far-infrared and infrared regions arises from two types of contributions, those from intraband and those from interband optical transitions. In the far-infrared region, the optical response is dominated by the intraband transition\cite{PhysRevLett.96.256802, RevModPhys.82.2673}. In the mid- to near-infrared region, the optical absorbance is attributable primarily to interband transitions. This response is nearly frequency independent and is equal to a universal value determined by the fine-structure constant $e^2/\hbar c$ in pristine graphene\cite{PhysRevLett.96.256802, Nair1308}. However, the optical absorption in graphene can be controlled through electrostatic gating, which shifts the Fermi energy and induces Pauli blocking of the optical transitions\cite{Wang206, Li2008}. 

When the Fermi level of graphene is shifted away from the charge neutral point, the absorption in the far-infrared region gets enhanced (compared to the universal absorption) and gets diminished in the infrared region as in figure \ref{terahertz_exp}(c). The amount of enhancement (yellow part in figure~\ref{terahertz_exp}(b)) equals to the amount of diminishment (green part in figure~\ref{terahertz_exp}(b)). A common argument is that the rise in the far-infrared region is due to phonon-assisted intraband processes made possible by the empty levels made available by doping, and the dip in the infrared region is due to the missing phononless vertical interband processes caused by the doping\cite{horng2011drude}. The equivalence of the area  of the far infrared  rise and infrared dip  is attributed to a   sum rule. There are several problems however: If phonon-assisted absorption played an important role in the far-infrared region, it should make an important contribution in the infrared region. The sum rule   applies to similar physical processes, and should not be used  to explain the intensities of two different processes.

%I'm thinking of the rise of far infrared region absorption might be related to the coherence length of electron in graphene. In Figure \ref{infra_far_infra_red_absorption}, the intraband absorption in the far infrared region mostly lies below $300 cm^{-1}$, which is about $0.037eV$. The electronic dispersion in the linear region approximately takes the form of 
%\begin{equation}
%E(\boldsymbol{q})=\frac{3}{2}ta\mid\boldsymbol{q}\mid, t=2.7eV,a=1.42\textup{\AA}
%\end{equation}
%where $\boldsymbol{q}$ is the momentum measured relative to the Dirac points. Then when $\Delta E=0.037eV$, the extra momentum needed for an intraband process is about $\mid\boldsymbol{q}\mid=6.4\times 10^5 cm^{-1}$. The wavelength of electron with momentum $\mid\boldsymbol{q}\mid$ is $\lambda = \frac{1}{\mid\boldsymbol{q}\mid}=15.6nm$. In the far infrared region, this number could even be bigger. This is in the similar order of magnitude as the electron coherence length in graphene. 

%Once the coherence length region is reached, the intraband process won't require strict momentum conservation. Thus phonon is not required for such absorption, that the rise in far infrared region and dip in infrared region are both phononless process. Then sum rule could hold.\\

%\begin{figure}[h!]
%\includegraphics[width=12cm]{sum_rule_diffEf}
%\caption{Far infrared and infrared region absorption with different biases}
%\label{sum_rule_diffEf}
%\end{figure}

Figure~\ref{terahertz_exp}(d) shows both the rise in the far-infrared region and the  compensating dip in the infrared region  as the bias voltage changes. The mild asymmetry between the hole response and electron response on either side of 0 bias   is due to asymmetric  carrier response when the bias is  applied in different directions, even though the Dirac cone is quite symmetric in the region we are interested in\cite{doi:10.1021/nl301496r}. The absorption enhancement   equals the absorption diminishment suggests a sum rule which should be sought relating to the same physical process. The infrared dip is clearly due to missing phononless interband electronic transitions when the Fermi level is shifted, and we must look to the far-infrared region  for added phononless electronic transitions.

\begin{figure}[h]
\includegraphics[width=9cm]{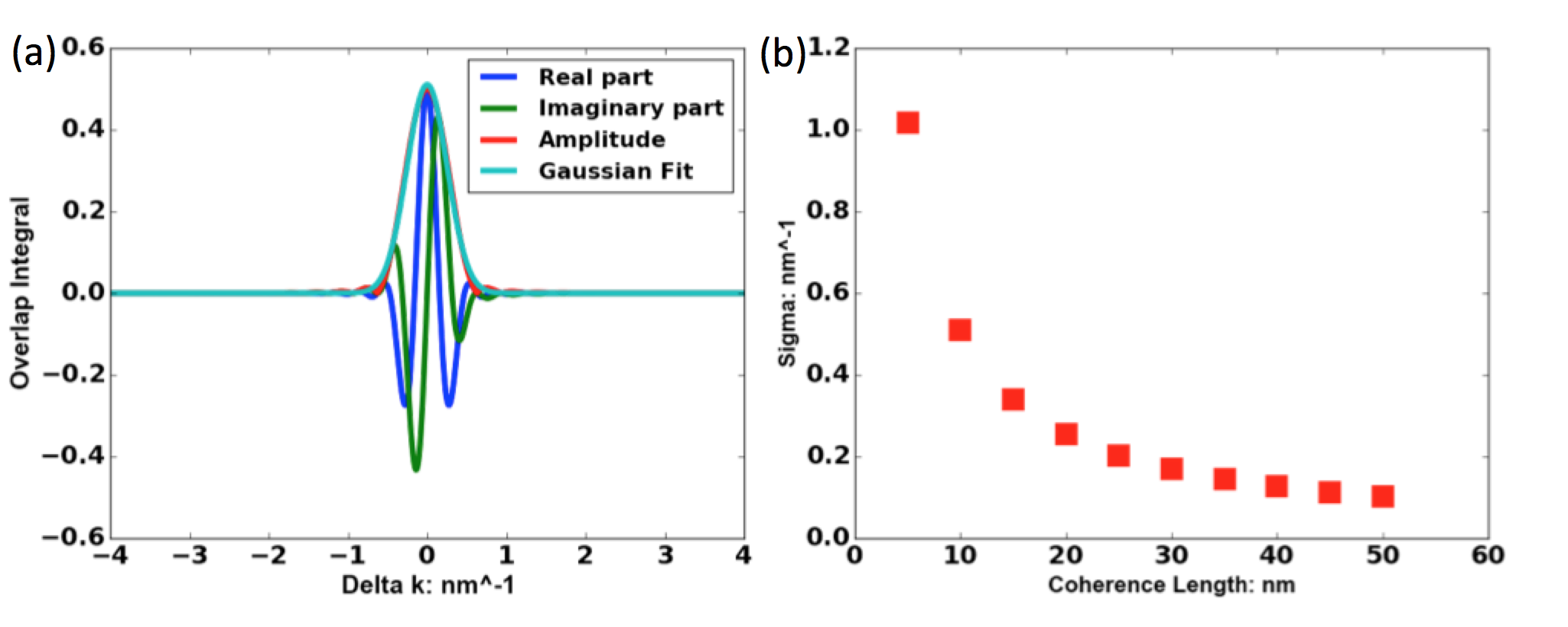}
\caption{(a) Overlap integral of two plane waves with momentum different by $\Delta k$: Assuming the coherence length $l$ is $20 nm$, and the integral is in the range of $0$ to $20nm$. A $sin(\frac{\pi x}{l})$ decay is added to the plane wave, which guarantees the wave function goes to 0 at the boundary smoothly. The blue line is the real part of the integral, the green line is the imaginary part, and the red line is the amplitude of the integral. We fit the amplitude by a Gaussian function(light blue line), and the Gaussian is centered at $0$ and has standard deviation $0.25nm^{-1}$. (b) Fit the overlap integral amplitude with a Gaussian function, the standard deviation of the gaussian goes down as the coherence length goes up.}
\label{overlap_integral}
\end{figure}

Are momentum conserving phononless intraband electronic absorption transitions  possible? A sketch on a band diagram would suggest not (figure~\ref{terahertz_exp}(a)). The intraband process lies at very low energy absorption region, which means that the transition is between two electronic states with very close wavevectors, if we consider the Bloch state in the graphene band structure. The overlap between electronic states with different wavevectors would be zero if the electron had infinite coherence length and the sample was infinitely large. However, the coherence length of electron could not be infinite in practice, and is on the order of ten nanometers\cite{doi:10.1021/nl104134a, PhysRevB.86.155403,doi:10.1021/nl8032697}. This makes the overlap between electronic states with different wavevectors not exactly zero. When two wavevectors are very close to each other, the overlap becomes significant. The major difference between two Bloch states with close momentums lies in the Bloch modulation, which is essentially plane waves with different momentum $k_1,k_2$. Figure~\ref{overlap_integral}(a) shows the overlap integral of two plane waves with momentum different by $\Delta k$. The blue line is the real part of the integral, the green line is the imaginary part, and the red line is the amplitude of the integral. The light blue line is  a Gaussian function fitting to the amplitude curve. Figure~\ref{overlap_integral}(b) shows the standard deviation of the Gaussian function goes down as the coherence length goes up. 

The Gaussian-like behavior of the overlap of two Bloch states implies that pure electronic transition could take place even though two Bloch  states have slightly different momenta.  The slightly nonvertical  or ``wobbling" effect of  the electronic transitions is enabled                                                                                                                                                                                                                                                                                                                                                                                                                                                                                                                                                                                                                                                                                                                                                                                                                                                                                                                                                                                                                                                                                                                                                   by  the finite coherence length of the  electrons, causing a coupling and broadening of the Bloch waves, which are now imperfect enforcers of momentum conservation.  The momentum conservation constraint in the Bloch basis becomes a Gaussian instead of a Dirac delta function, given as Eq (\ref{momentum_conservation_tolerance}).

\begin{equation}
\label{momentum_conservation_tolerance}
g(\Delta k_x,\Delta k_y)=\frac{1}{\sqrt{2\pi}\sigma}e^{-\frac{(\Delta k_x)^2+(\Delta k_y)^2}{2\sigma^2}},
\end{equation}
implying  that  both  vertical or close to vertical processes are allowed to happen in phononless absorption.  $\sigma$ controls the strictness of the momentum conservation,  becoming a Dirac delta function at infinite electron coherence length $\sigma\to 0$. According to figure~\ref{overlap_integral}, we choose the standard deviation for momentum as $0.2nm^{-1}$. The electronic dispersion around the Dirac cone follows $E(\bm{q})=3/2ta\abs{\bm{q}}, t=2.7eV, a=0.142nm$\cite{RevModPhys.81.109}. The standard deviation for momentum tolerance in terms of energy becomes $0.11eV^{-1}$. %The sum rule implied in the model won't be affected by the exact value of $\sigma$. 

Except for the Bloch modulation, the pure electronic transition matrix elements change only slightly at different energies. Thus we model the transition intensity between two states by considering their Bloch modulation overlap,  approximated by a Gaussian function as we see qualitatively in figure~\ref{overlap_integral}. In the relevant region   the electronic dispersion is quite linear, as was used used in the model. Assuming the Fermi level is shifted down from the Dirac cone by $E_f$, then the absorption at incident light energy $E$ can be expressed as

\begin{equation}
\label{absorption_phononless}
\begin{split}
\sigma(E)=&\frac{1}{E}\int_0^{2\pi}d\theta_1\int_0^{2\pi}d\theta_2\int_0^{E}de(E_F+e)\abs{E_F-(E-e)}\\
&g((E_F+e)\cos\theta_1-\abs{E_F-(E-e)} \cos\theta_2, \\
&(E_F+e)\sin\theta_1-\abs{E_F-(E-e)} \sin\theta_2)
\end{split}
\end{equation}

We set the energy of the center of Dirac cone to  0. The excited electron   has energy $-(E_F+e)$, and the density of electronic states at energy $-(E_F+e)$ is proportional to $E_F+e$. The final state has energy $-(E_F-(E-e))$, and the density of states at the final state with energy is proportional to $\abs{E_F-(E-e)}$. The initial state has momentum proportional to $((E_F+e)\cos\theta_1,(E_F+e)\sin\theta_1)$, and the final state has momentum proportional $(\abs{E_F-(E-e)} \cos\theta_2, \abs{E_F-(E-e)} \sin\theta_2)$. This formula holds if the Fermi level is shifted up by $E_F$

\begin{figure}[h]
\includegraphics[width=9cm]{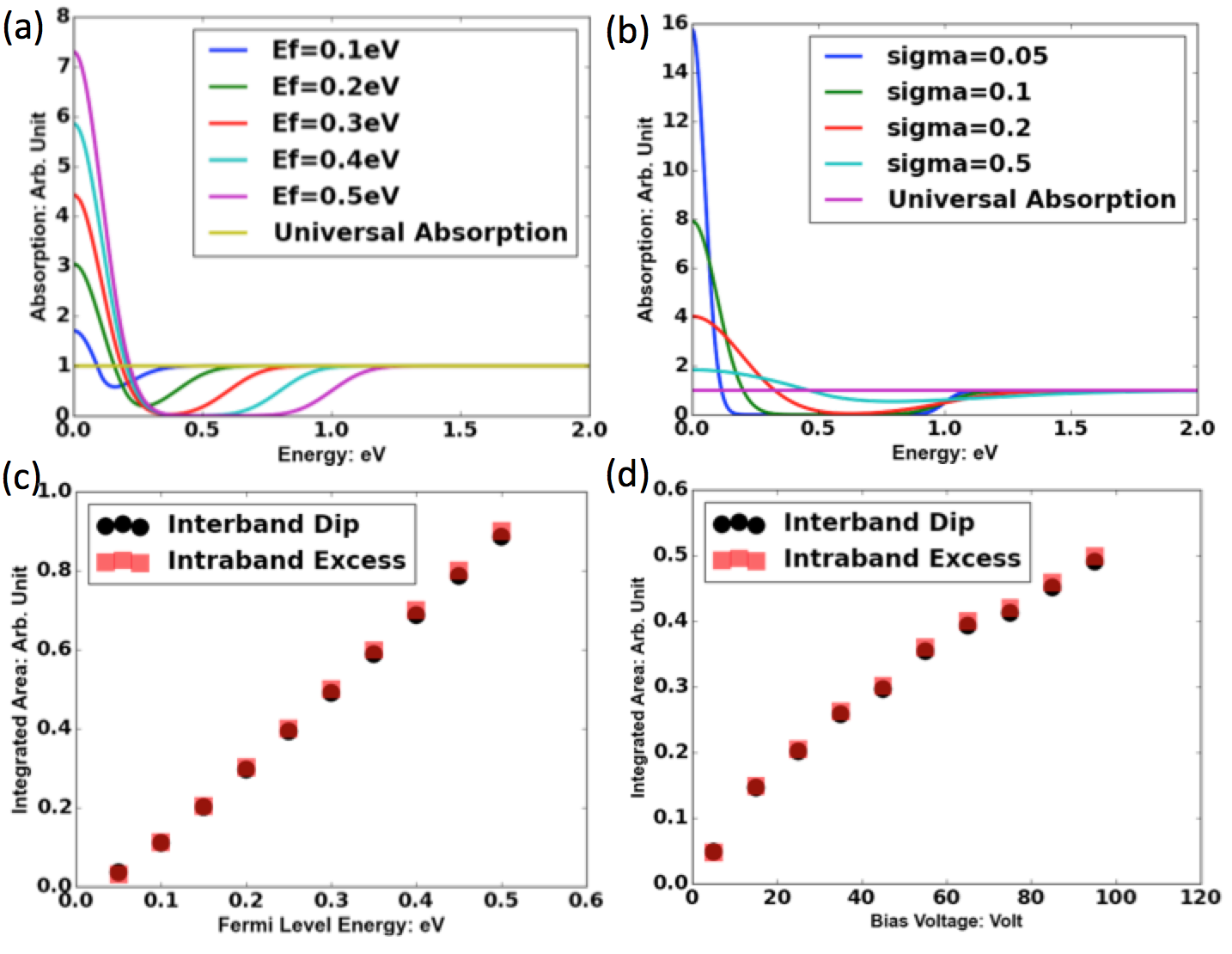}
\caption{Graphene far infrared and infrared region absorption:  (a) is the simulated absorption results according to Eq (\ref{absorption_phononless}) when the Fermi level is shifted away from cone center by 0.1, 0.2, 0.3, 0.4 and 0.5eV. The plot shows the dip of absorption in the infrared region and the rise in the far infrared region. The universal absorption is also shown. (b) shows how the absorption curve changes as we choose different $\sigma$'s, assuming the Fermi level is shifted away from the cone center by 0.5eV. (c) shows the integrated values of the enhanced absorption in the far-infrared region and the diminished absorption in the infrared region as a function of the amount that the Fermi level shifted by. The enhancement compensates the diminishment. The integrated area almost increase linearly in terms of the shift amount of the Fermi level. (d) shows the integrated values of the enhanced absorption in the far-infrared region and the diminished absorption in the infrared region as a function of the gating voltage. The trend agrees well with the experiment result in figure~\ref{terahertz_exp}(d). }
\label{terahertz_simulation}
\end{figure}

Figure~\ref{terahertz_simulation} shows results calculated based on Eq(\ref{absorption_phononless}).  Figures~\ref{terahertz_simulation}(a) reveal the rise of absorption in the far infrared region and the dip in the infrared region when the Fermi level is shifted away from the cone center by different amounts.  Figure~\ref{terahertz_simulation}(c) shows the corresponding area, revealing the sum rule. These areas linearly increase with   the  Fermi level shift,  because the number of phononless electronic transition processes is proportional to the density of states of electron. The area of the rise is proportional to the number of extra processes and the area of the dip is proportional to the number of lost processes due to the applied bias. 

The amount that the Fermi level shift  is roughly the square root of the bias voltage\cite{Wang206, doi:10.1021/nl901572a, doi:10.1021/nl301496r}. Using a simple capacitor model,   $\mid E_F(\Delta V_g)\mid = \hbar v_F\sqrt{\pi\mid\alpha_0(\Delta V_g)\mid}$, where $v_F=1\times10^6m/s$ is the Fermi velocity of Dirac fermions in graphene, and  $\alpha_0\approx7\times10^{10}cm^{-2}V^{-1}$ is the gate capacitance in electron charge\cite{Wang206}. Figure~\ref{terahertz_simulation}(d) shows the integrated values of the enhanced absorption in the far-infrared region and the diminishment absorption in the infrared region as a function of the gating voltage. The trend agrees well with the experimental result in figure~\ref{terahertz_exp}(d).

Figure~\ref{terahertz_simulation}(b) shows the changes of the absorption curve as different $\sigma$'s are used for a fixed Fermi level shift 0.5eV. The far infrared absorption gets a sharper raise and the infrared absorption gets a more significant drop when a smaller $\sigma$ is used. The sum rule implied in the model is not affected by the exact value of $\sigma$.

It is clear from our model that the rise in the far-infrared region is due to extra intraband pure electronic transitions when the Fermi level is shifted and the dip in the infrared region is due to the lack of initial electronic state which reduces some electronic transitions. This should not be a special phenomenon in graphene, we should also be able to observe similar effects in materials such as doped semiconductors with small enough band gaps.  

Low temperature transport experiments have shown that the coherence length of charge carriers in graphene is proportional to $1/T^{\frac{1}{2}}$\cite{PhysRevLett.97.016801}.  A lower temperature implies a longer electron coherence length and thus a smaller momentum tolerance according to our model. So we expect to observe a sharper far infrared absorption enhancement at a lower temperature as implied in figure~\ref{terahertz_simulation}(b). 

\section{Conclusion}

We developed a quantum description for the intraband absorption and interband absorption in graphene. Under the framework both absorption processes are caused by the same physical mechanism, pure electronic transition between quantum states. Thus they can obey a sum rule.  We have explained the rise of far infrared absorption and the dip of infrared absorption in a doped graphene,  and their dependence on the gate bias. Our model naturally lead to results obeying the sum rule.

This work was supported by the STC Center for Integrated Quantum Materials, NSF Grant No. DMR-1231319. The authors thank professor Efthimios Kaxiras, Wei Chen, Shiang Fang, Wenbo Fu, Ping Gao, Lu Shen for helpful discussions.

%\begin{acknowledgments}
%We wish to acknowledge the support of the author community in using
%REV\TeX{}, offering suggestions and encouragement, testing new versions,
%\end{acknowledgments}

%\appendix

%\nocite{*}

\bibliography{graphene_terahertz}% Produces the bibliography via BibTeX.

\end{document}